\documentclass{iopart}

\usepackage{graphicx}
\usepackage{array}
\usepackage{iopams} 
\usepackage{bm}
\usepackage[english]{babel}

\usepackage[latin1]{inputenc}
\begin{document}

\title[Auxiliary field method for exponential potentials]{Auxiliary field method and analytical solutions of the Schr\"{o}dinger equation with exponential potentials}

\author{Bernard Silvestre-Brac$^1$, Claude Semay$^2$ and Fabien Buisseret$^2$}

\address{$^1$ LPSC Universit\'{e} Joseph Fourier, Grenoble 1,
CNRS/IN2P3, Institut Polytechnique de Grenoble, 
Avenue des Martyrs 53, F-38026 Grenoble-Cedex, France}
\address{$^2$ Groupe de Physique Nucl\'{e}aire Th\'{e}orique, Universit\'{e}
de Mons-Hainaut, Acad\'{e}mie universitaire Wallonie-Bruxelles, Place du Parc 20,
B-7000 Mons, Belgium}
\eads{\mailto{silvestre@lpsc.in2p3.fr}, \mailto{claude.semay@umh.ac.be}, 
\mailto{fabien.buisseret@umh.ac.be}} 
\date{\today}

\begin{abstract}
The auxiliary field method is a new and efficient way to compute approximate analytical eigenenergies and eigenvectors of the Schr\"{o}dinger equation. This method has already been successfully applied to the case of central potentials of power-law and logarithmic forms. In the present work, we show that the Schr\"{o}dinger equation with exponential potentials of the form $-\alpha\, r^\lambda\, \exp(-\beta r)$ can also be analytically solved by using the auxiliary field method. Formulae giving the critical heights and the energy levels of these potentials are presented. Special attention is drawn on the Yukawa potential and the pure exponential one. 
\end{abstract}

\submitto{J. Phys. A: Math. Theor.}
\maketitle

\section{Introduction}

Much work has been devoted to the computation of analytical solutions of the Schr\"{o}dinger equation for the early years of quantum mechanics, especially when the Hamiltonian of the problem admits bound states. Indeed, it is often useful in physics to have a guess about the behavior of the eigenenergies of a system in terms of the various parameters of the Hamiltonian. This can be particularly relevant when one tries to fit the parameters of a model to some experimental data. 

There are many methods allowing to find approximate analytical solutions of the Schr\"{o}dinger equation for bound state problems: WKB method, variational method, etc. (see \cite{flu} for example). In previous works \cite{af,af2}, we have proposed and extensively studied a new method to compute the eigenenergies and wave functions of a given Hamiltonian admitting bound states. This method is based on auxiliary fields (also known as einbein fields \cite{af1}) and can lead to analytical approximate results. Its general principle will be recalled in Section~\ref{grmeth}. Up to now, the auxiliary field method (AFM) has been mainly applied in the case of central potentials of the form $a r^\lambda+br^\eta$. We have shown, in particular, that the qualitative features of the exact energy spectra are well reproduced and that the analytical formulae can be greatly improved by a comparison with a numerical resolution of the problem. Then they become very accurate \cite{af,af2}. 

In the present work, we intend to apply the AFM to find approximate analytical formulae for central potentials of exponential form, that is $-\alpha\, r^{\lambda}\, \rme^{-\beta\, r}$. We thus start with the following Hamiltonian
\begin{equation}\label{h0def}
	H=\frac{\vec p^{\, 2}}{2m}-\alpha\, r^\lambda\, \rme^{-\beta\, r},
\end{equation}
where $\alpha$ and $\beta$ are both positive and nonzero real parameters. It is interesting to define the new variables $\{\vec x=\beta\, \vec r,\ \vec q=\vec p/\beta\}$ and to work with the dimensionless Hamiltonian ${\cal H}=2mH/\beta^2$ instead of $H$. Indeed, one has then
\begin{equation}\label{hdef}
	{\cal H}=\vec q^{\, 2}-g\, x^\lambda\, \rme^{-x},
\end{equation}
 with
\begin{equation}\label{gdef}
	g=\frac{2m\alpha}{\beta^{\lambda+2}}.
\end{equation}
The rewriting of the problem under this particular form emphasizes the fact that the dimensionless parameter $g$ given by (\ref{gdef}) contains all the relevant information of $H$. Only the eigenenergies $\epsilon(g)$ of~(\ref{hdef}) have thus to be computed; the ``physical" eigenenergies will then be given by $E(m,\alpha,\beta)=\beta^2\, \epsilon(g)/(2m)$, a relation which expresses the scaling properties of the Schr\"{o}dinger equation.

We first show in section~\ref{expp} that the pure exponential potential ($\lambda=0$) can be analytically solved using the AFM provided that the energy spectrum of the Schr\"{o}dinger equation with a linearly rising potential is analytically known. The pure exponential potential is common in hadronic physics for example, when a screened confining potential is considered \cite{lucha,gonz}. Then, we show that some general results can be obtained for $\lambda\neq0$ in section~\ref{genp}, and write an explicit energy formula for the Yukawa potential ($\lambda=-1$) in section~\ref{yuka}. The importance of this potential in theoretical physics is such that it also deserves a specific study. We then present improved energy formulae for the pure exponential and the Yukawa potentials in section~\ref{impfore}.

An interesting feature of exponential potentials is that they all admit a \textit{finite} number of bound states that depends on the dimensionless parameter $g$, ruling the potential depth, and defined by (\ref{gdef}). There exists thus ``critical heights": Potential depths beyond which new bound states appear. We refer the reader to \cite{brau1} for detailed explanations about how to compute critical heights in a given potential. Accurate formulae giving these critical heights are given for the pure exponential and the Yukawa potentials. Finally, we sum up our results in section~\ref{conclu}. The definition and the properties of the Lambert function, that will frequently appear in our calculations, are given in \ref{lambprop}. Additional analytical approximations for the energy levels and some critical heights for the Yukawa potential are given in \ref{envyuk} and \ref{gyuk} respectively.

\section{Auxiliary field method}\label{grmeth}
\subsection{General principle}
We recall here the main points of the method that we presented in \cite{af}. The problem is to find an analytical approximate solution of the eigenequation $	H \left|\Psi\right\rangle=E\left|\Psi\right\rangle$, $H$ being a Hamiltonian of standard radial form 
\begin{equation}\label{inco}
	H=T(\vec p^{\, 2})+V(r).
\end{equation}

The procedure is the following. Let us assume that $H_A=T(\vec p^{\, 2})+\nu\, P(r)$ admits bound states with an analytical spectrum, $\nu$ being a real parameter. We thus know the solution of  
\begin{equation}\label{analy}
	H_A \left|\Psi_A(\nu)\right\rangle=E_A(\nu)\left|\Psi_A(\nu)\right\rangle.
\end{equation}
Let us define the function $I(x)=K^{-1}(x)$ and the operator $\hat \nu$ from the relation
\begin{equation}\label{rdef}
	\hat\nu\equiv\frac{\partial_rV(r)}{\partial_rP(r)}= K(r).
\end{equation}
The spectrum of the following Hamiltonian 
\begin{equation}\label{htdef}
	\tilde H(\nu)=T(\vec p^{\, 2})+\nu\, P(r)+V\left[I(\nu)\right]-\nu\, P\left[I(\nu)\right], 
\end{equation}
is then analytically known provided that $\nu$ -- the auxiliary field -- is assumed to be a real parameter. Using (\ref{analy}), we have indeed
\begin{equation}\label{en1}
	E(\nu)=E_A(\nu)+V\left[I(\nu)\right]-\nu\, P\left[I(\nu)\right].
\end{equation}
The physical spectrum is finally given by the eigenstates $\left|\Psi_A(\nu_0)\right\rangle$ and the eigenenergies $E(\nu_0)$, with $\nu_0$ minimizing the total energy~(\ref{en1}), \emph{i.e.} satisfying
\begin{equation}\label{enmin}
\left.\frac{\partial E(\nu)}{\partial\nu}\right|_{\nu=\nu_0}=0.
\end{equation}

This method can be justified as follows. The arbitrary potential $V(r)$ is replaced by $\tilde V(r,\nu)=\nu\, P(r)+V\left[ I(\nu)\right]-\nu\, P\left[I(\nu)\right]$, a function of the auxiliary field $\nu$ and of an analytically solvable potential $P(r)$. The key point is that the auxiliary field should rigorously be eliminated as an operator from Hamiltonian~(\ref{htdef}). When doing this, one is led to $\left.\delta_\nu \tilde V(r,\nu)\right|_{\nu=\hat\nu}=0\Rightarrow \hat\nu=K(r)$, and to $\tilde V(r,\hat\nu)=V(r)$. This means that Hamiltonians~(\ref{inco}) and (\ref{htdef}) are equivalent up to an elimination of the auxiliary field as an operator, the auxiliary field being then defined by (\ref{rdef}). The AFM would consequently lead to the exact results if the real number $\nu$ were replaced by the operator $\hat \nu$. As we use the real number $\nu$ in order to have analytical solutions, the results are approximate. One has then $\nu_0\approx \left\langle \Psi_A(\nu_0)\right|\hat\nu\left|\Psi_A(\nu_0)\right\rangle$, with $\hat \nu$ and $\nu_0$ given by (\ref{rdef}) and (\ref{enmin}) respectively \cite{af}. Our method can consequently be regarded as a ``mean field approximation" with respect to a particular auxiliary field which is introduced to simplify the calculations.

Let us notice that the method we have presented is not variational, and that it is exact when $V(r)=P(r)$, with trivially $\hat\nu=1$. The main technical problem is the determination of analytical solutions for (\ref{rdef}) and (\ref{enmin}). Such a task has already been fulfilled in a satisfactory way for many interesting and nontrivial potentials \cite{af2}. As we will see in the latter, it can also be achieved for the Yukawa and the pure exponential potentials. 

\subsection{Useful formulae}

We have shown in \cite{af} that the eigenvalues of the Hamiltonian 
\begin{equation}
\label{eq:plpot}
H_{\eta}(a) = \frac{\vec{p}^{\, 2}}{2m}+ a\, \textrm{sgn}(\eta)
r^\eta
\end{equation}
can be written under the form
\begin{equation}
\label{eq:eigenerplpot}
e_\eta(a)=\frac{2+\eta}{2 \eta} (a |\eta|)^{2/(\eta+2)}
\left ( \frac{N_\eta^2}{m} \right )^{\eta/(\eta+2)},
\end{equation}
where $N_\eta$ depends on the radial $n$ and orbital $l$ quantum numbers, as well as on $\eta$ \emph{a priori}. An interesting feature of the AFM is that the functional form of (\ref{eq:eigenerplpot}) does not depend on the chosen potential $P(r)$, provided that this potential is of power-law form (typically one can choose $r^2$ or $-1/r$) \cite{af2}. $N_\eta$ depends only on the particular form of $P(r)$: One logically obtains $N_\eta=2n+l+3/2$ or $n+l+1$ according to the choice $P(r)=r^2$ or $-1/r$. Since the form of $N_\eta$ is an artifact of the AFM, it can be modified to better fit the exact results. In particular, we have shown that, for any physical values of $\eta$ ($\eta > -2$), a good form for $N_\eta$ is given by
\begin{equation}
\label{eq:grandNl}
N_\eta = b(\eta) n + l + c(\eta).
\end{equation}
Simple expressions for the functions $b(\eta)$ and $c(\eta)$ have been proposed in \cite{af} in order to give an approximation as precise as $10^{-3}$ for the most interesting (the lowest) values of the quantum numbers $n$ and $l$. For example,
\begin{equation}\label{bcdef}
	b(\eta)=\frac{41\eta+86}{13\eta+58},\quad c(\eta)=\frac{5\eta+17}{2\eta+14},
\end{equation}
leads to very accurate energy formulae \cite{af}. Finally, it is worth mentioning that the energy formula~(\ref{eq:eigenerplpot}) together with (\ref{eq:grandNl}) and (\ref{bcdef}) gives the exact result in the following important cases: 
\begin{itemize}
\item the harmonic oscillator, for which $N_{2}=2n+l+3/2$;
\item the Coulomb potential, for which $N_{-1}=n+l+1$.
\end{itemize}

As we have analytical and accurate energy formulae for any power-law potential at our disposal, we will assume in this work that $P^{(\eta)}(r)=\textrm{sgn}(\eta)\, r^\eta$ is a good starting point to apply the AFM. Setting $P(r)=P^{(\eta)}(r)$, we have shown in \cite{af2} that the energy spectrum of Hamiltonian
\begin{equation}
	H= \frac{\vec{p}^{\, 2}}{2m}+ V(r)
\end{equation}
is given by
\begin{equation}
\label{eq:enu0frompl}
E(\nu_0)=\frac{|\eta|}{2} \nu_0 I(\nu_0)^\eta + V(I(\nu_0)),
\end{equation}
the optimal value $\nu_0$ being determined from the equation
\begin{equation}
\label{eq:nu0frompl}
|\eta|\nu_0 I(\nu_0)^{\eta+2}=\frac{N_\eta^2}{m}
\end{equation}
and the function $I(\nu)$ coming from the relation
\begin{equation}
\label{eq:Jnufrompl}
|\eta| \nu I(\nu)^{\eta - 1}=V'(I(\nu)).
\end{equation}
$I(\nu)$ depends only on the potential $V(r)$ and not on a particular eigenstate, while $\nu_0$ depends both on the potential (through the $I$ function) and on the state under consideration (through $N_\eta$). Equations~(\ref{eq:enu0frompl})-(\ref{eq:Jnufrompl}) are general results of the AFM, that will be used intensively in this work. 

We have shown in \cite{af2} that replacing the power $\eta$ by $\lambda$ does not change the functional form of $I(\nu)$ but substitutes simply the number $N_\eta$ by $N_\lambda$. The AFM cannot say what is the optimal value for the power of $P(r)$. It is necessary to resort to a numerical analysis. Moreover it was also shown in \cite{af2} that the approximate eigenvalues are ruled by the correct scaling laws.

\section{Pure exponential potential}\label{expp}
\subsection{Energy spectrum}
The Schr\"{o}dinger equation with pure exponential potential corresponds to the Hamiltonian~(\ref{hdef}) in which $\lambda=0$, \emph{i.e.}
\begin{equation}\label{hdefe}
	{\cal H}_e=\vec q^{\, 2}-g\, \rme^{-x}.
\end{equation} 
The negative eigenenergies $\epsilon(g)$ of such a Hamiltonian can be analytically computed for $l=0$ only. But even in this case, the expression of $\epsilon(g)$ is not very tractable since it is formally defined by the relation \cite{flu}
\begin{equation}\label{enexa}
J_{2\sqrt{-\epsilon(g)}}(2\sqrt g)=0,	
\end{equation}
where $J_\rho$ is a Bessel function of the first kind. Consequently, it is of interest to find an analytical energy formula for the energy levels of the pure exponential potential that is of simpler use than (\ref{enexa}) for $l=0$ and, above all, that remains valid for arbitrary $n$ and $l$ quantum numbers.  

Equations~(\ref{eq:enu0frompl})-(\ref{eq:Jnufrompl}) can be particularised in the present case. Obviously one has to set $V(x)=-g\, \rme^{-x}$ and $m=1/2$, while a particularly convenient choice is $P(x)=P^{(1)}(x)=x$. A quick calculation leads to 
\begin{equation}
	I(\nu)=-\ln\left(\frac{\nu}{g}\right)
\end{equation}
and to the energy formula
\begin{equation}\label{enexp}
	\epsilon_e(\nu_0)=-\nu_0-\frac{\nu_0}{2}\ln\left(\frac{\nu_0}{g}\right),
\end{equation}
where $\nu_0$ satisfies the equation
\begin{equation}\label{eqpar1}
	-\nu_0\left[\ln\left(\frac{\nu_0}{g}\right)\right]^3=2N^2.
\end{equation}
Notice that, in principle, $N= N_1$ in this last equality. But, as mentioned above, another choice $N_\lambda$ with an arbitrary value of $\lambda$ is \emph{a priori} possible. It is worth noting that, from (\ref{rdef}), we have $K(x)=g\, \rme^{-x}$ and thus $\nu_0\in[0,g]$. 

By defining 
\begin{equation}
\label{defZ}
	x_0=\left(\frac{\nu_0}{g}\right)^{1/3}\quad {\rm and}\quad Z=\frac{1}{3}\left(\frac{2N^2}{g}\right)^{1/3},
\end{equation}
we can recast (\ref{eqpar1}) into the form 
\begin{equation}\label{eqelim}
	-x_0\, \ln x_0=Z.
\end{equation}
$x_0\in[0,1]$ by definition, and the function $F(x_0)=-x_0 \ln x_0$ is such that $F(0)=F(1)=0$, with a maximum in $F(1/\rme)=1/\rme$. It means that (\ref{eqelim}) has no solution if $Z>1/\rme$; the solution of (\ref{eqelim}) turns out to be
\begin{equation}\label{thesolu}
	x_0(Z)=\rme^{W_0(-Z)}=-\frac{Z}{W_0(-Z)} \quad {\rm with} \quad 0< Z\leq1/{\rm e},
\end{equation}
where $W_0$ is the principal branch of the Lambert function, whose main properties are given in \ref{lambprop}. The choice of $W_0$ ensures that the total energy has the correct limit for small $\beta$ (see below). Equation~(\ref{enexp}) now becomes
\begin{equation}
\label{epsnu0}
	\epsilon_e(\nu_0)=\epsilon_e(g x_0(Z)^3)=-g\, \rme^{3 W_0(-Z)}\, \left[1+\frac{3}{2}\, W_0(-Z)\right],
\end{equation}
and the energy, rewritten in physical units, is finally given by
\begin{equation}\label{enfinal}
\fl	E_e=\frac{\beta^2}{2m}\epsilon_e=-\alpha\, \rme^{3 W_0(-Z)}\, \left[1+\frac{3}{2}\, W_0(-Z)\right],\quad {\rm with}\quad Z=\frac{1}{3}\left(\frac{N^2 \beta^2}{m \alpha}\right)^{1/3}.
\end{equation}
The limit $\beta\to 0$ is interesting. Indeed, $-\alpha\, \rme^{-\beta r}\approx-\alpha+\alpha\beta r$ in this case: The potential is nearly a linear confining one. Recalling that $g=2m\alpha/\beta^2$ for the pure exponential potential, it can be checked that, at small $\beta$, 
\begin{equation}
	E_e=-\alpha+\frac{3}{2}\, m^{-1/3}\, \left[\alpha\beta N\right]^{2/3}+O(\beta^{4/3}).
\end{equation}
This last formula is exactly the one which is expected in the case of a linear potential [see (\ref{eq:eigenerplpot}) for $a=\alpha\beta$ and $\lambda=1$].

For computational reasons, the Lambert function may be uneasy to deal with. We mention that we can use the approximate expression
\begin{equation}
\fl	\rme^{3 W_0(-Z)}\left[1 + \frac{3}{2} W_0(-Z)\right]\approx\frac{\rme^{-3Z}}{100}\left[100-150\, Z - 580\, Z^2 + 524\, Z^3\right] 
\end{equation}
which leads to an absolute error below 0.008 for $0 \leq Z \leq 1/\rme$. 

We finally point out that the energy spectrum of potentials like $\left(a-b\, \rme^{-cr}\right)$ are trivially obtained by adding a constant term to formula (\ref{enfinal}). Such potentials are of interest in hadronic physics when one considers a screened confining potential to take into account possible string-breaking effects \cite{lucha,gonz}.  

\subsection{Critical heights}\label{gcexp}

Let us consider first the AFM framework. By definition, the critical heights $g_{nl}$ of the exponential are such that $\epsilon_e(g=g_{nl})=0$, with $\epsilon_e$ given by (\ref{epsnu0}). According to this last equation, one finds that the energy vanishes for $Z=2\, \rme^{-2/3}/3$, a value which is lower than $1/\rme$ as required by (\ref{eqelim}). Equivalently, one can say that the critical heights are given by
\begin{equation}\label{gc1}
g_{e;nl}=\frac{\rme^2}{4}\, N^2_{nl}.	
\end{equation}
They are such that, if $g > g_{n_0l_0}$, the potential admits a bound state with the given quantum numbers $n_0$, $l_0$. It is remarkable that our approximation scheme, based on the linear potential $g x$ for which an infinite number of bound states is present, is able to predict that only a finite number of bound states will be present in the pure exponential potential. This is another test of the ability of the AFM to analytically reproduce the qualitative features of a given eigenvalue problem. Let us note that the AFM does not give any information about the optimal power-law potential to determine the form of the number $N_{nl}$.

Equation~(\ref{enexa}) can also lead to a determination of the exact (as coming from the Schr\"{o}dinger equation) critical heights $g_{n0}$; let us denote them as $g^*_{n0}$. Indeed these critical heights are such that they correspond to a energy level $\epsilon(g^*_{n0})=0$.  We are thus led to the equation
\begin{equation}
	J_0(2\sqrt{ g^*_{n0}})=0\Rightarrow  g^*_{n0}=\frac{j^2_n}{4},
\end{equation}
where $j_n$ is the $(n+1)^{{\rm th}}$ zero of the Bessel function $J_0$. At large $n$, these are given by $j_n\approx \pi(n+3/4)$ \cite{Abra}, leading to the asymptotic expression
\begin{equation}
\label{gcbess}
	g^*_{n0}\approx\frac{\pi^2}{4}\, \left(n+\frac{3}{4}\right)^2.
\end{equation}
This result is qualitatively similar to (\ref{gc1}), stating that $g_{n0}\propto N_{n0}^2$.  

We can try to use formula~(\ref{gcbess}) to improve the result~(\ref{gc1}). Assuming that $N_{nl}= b\, n+l+c$ with no constraint on parameters $b$ and $c$ (contrary to what is expected from (\ref{eq:grandNl})), it is easy to see that formulae~(\ref{gc1}) and (\ref{gcbess}) coincide for $g_{n0}$ with $b=\pi/\rme$ and $c=3\pi/(4\rme)$. We can then try a new relation to compute the critical heights 
\begin{equation}\label{gc1b}
g_{e;nl}=\left(\frac{\pi}{2} n+\frac{\rme}{2} l+\frac{3\pi}{8}\right)^2.	
\end{equation}
This formula is in good agreement with exact results: For $n\in[0,5]$ and $l\in[0,5]$, the minimal, maximal and mean relative errors are respectively 0.03\%, 8.7\% and 3.5\%. It is not possible to obtain so good agreement by choosing $b$ and $c$ values coming from (\ref{bcdef}). Some exact $g^*_{e;nl}$ values are given in table~\ref{tab2}; they are computed numerically by solving the Schr\"{o}dinger equation with a vanishing energy.

\begin{table}[ht]
\caption{\label{tab2}Some exact critical heights $g^*_{e;nl}$ for the exponential potential $-g\, \rme^{-x}$.}
\begin{indented}
\item[]\begin{tabular}{@{}ccccc}
\br 
$n$ & $l=0$ & 1 & 2 & 3\\
\mr
0& 1.446 & 7.049 & 16.313 & 29.258\\
1& 7.618 & 16.921& 29.880& 46.518\\
2& 18.722 & 31.526& 48.077& 68.346\\
3& 34.760 &50.948 & 71.002& 94.838 \\
\br
\end{tabular}
\end{indented}
\end{table}

Formula~(\ref{gc1b}) is fully analytical and is designed in order to reproduce correctly the behaviour of the exact critical heights for large $n$ and $l=0$, $g_{e;n0} \approx g^*_{n0}$. Unfortunately, the analytical asymptotic expression of $g^*_{0l}$ is not known, and an improvement of formula~(\ref{gc1b}) must rely on numerical arguments, in particular using the numerical values of $g^*_{e;nl}$, as given in table~\ref{tab2}. Moreover, a good overall fit formula needs to reproduce correctly not only the behaviour in terms of $n$ or $l$ separately, but must try to get the good behaviour in terms of $n$ and $l$ simultaneously. Since we want to stick as much as possible to the philosophy of AFM, we maintain the quadratic form  $g_{e;nl}\propto N^2$ and just add in the expression of $N$ a term proportional to $\sqrt{nl}$. Thus, we propose a new expression for the critical heights which reads
\begin{equation}
\label{gc1c}
g_{e;nl} \approx (1.566\,n+1.393\,l-0.125\sqrt{n\,l}+1.202)^2.
\end{equation}
The improvement as compared to (\ref{gc1b}) is very appreciable since, in this case, the minimal, maximal and mean relative errors are respectively 0.005\%, 4.5\% and 1.1\%.

Using (\ref{gc1}), one can write
\begin{equation}\label{zgc}
Z=\frac{2}{3}\rme^{-2/3}\left( \frac{g_{e;nl}}{g} \right)^{1/3}.
\end{equation}
This makes clear the dependence on the critical heights $g_{e;nl}$ for formula~(\ref{epsnu0}). It is then easy to see that $\epsilon_e=0$ when $g=g_{e;nl}$.

\section{The general case}
\label{genp}

In order to compute analytically the spectrum of Hamiltonian~(\ref{hdef}) for arbitrary values of $\lambda$, it is of interest to use the general result~(\ref{eq:eigenerplpot}), and consequently to choose $P(x)={\rm sgn}(\lambda)\, x^\lambda $. Equation~(\ref{rdef}) can be rewritten in the present case as
\begin{equation}\label{eqs}
\hat\nu=K(x)=g\frac{\rme^{-x}}{|\lambda|}(x-\lambda).
\end{equation}
The function $K(x)$ is such that $K(0)=-g\, {\rm sgn}(\lambda)$, $K(\lambda)=0$, and $\lim_{x\to \infty}K(x)=0^+$. It has a maximum in $K(\lambda+1)=g\rme^{-\lambda-1}/|\lambda|$, the position of this maximum being in the range $x\geq0$ for $\lambda\geq-1$. We can conclude from this discussion that $K(x)\in[0,g]$ if $-2<\lambda\leq-1$, $K(x)\in[0,g\, \rme^{-\lambda-1}/|\lambda|]$ if $-1<\lambda<0$ and $K(x)\in[-g,g\, \rme^{-\lambda-1}/|\lambda|]$ if $\lambda>0$. In the first case, the function $K(x)$ is monotonically decreasing while in the two last cases, it presents a maximum.

The inversion of (\ref{eqs}) can be achieved thanks to the Lambert function since $K(x)$ is of the form~(\ref{resol}). One finds 
\begin{equation}\label{ilsol}
	I_\lambda(\nu)=\lambda-W\left(-\rme^\lambda|\lambda|\frac{\nu}{g}\right),
\end{equation}
where one must be careful in the choice of either $W_0$ or $W_{-1}$, that will depend on the particular value of $\lambda$. Nevertheless, as both branches of the Lambert function share the same properties, it is possible to go a step further in our computations by using the generic solution~(\ref{ilsol}).

Let us first define
\begin{equation}
	x_0=\frac{|\lambda|\nu_0\rme^\lambda}{g}\quad {\rm and}\quad Y_\lambda=\frac{2N^2\rme^\lambda}{g},
\end{equation}
where $N= N_\lambda$ \emph{a priori}. But let us recall that another choice is possible as mentioned previously. Then, a rewriting of (\ref{eq:enu0frompl})-(\ref{eq:Jnufrompl}) with $V(x)=-g\, x^\lambda\, \rme^{-x}$, $m=1/2$, and $P(x)=P^{(\lambda)}(x)=-{\rm sgn}(x)\, x^\lambda$ leads to 
\begin{equation}\label{elneg}
	\epsilon_\lambda(x_0)=-\frac{g}{2\rme^\lambda}x_0^{\frac{2}{\lambda+2}}\, Y_\lambda^{\frac{\lambda}{\lambda+2}}\left[\frac{(\lambda+2)x_0^{\frac{1}{\lambda+2}}-Y_\lambda^{\frac{1}{\lambda+2}}}{Y_\lambda^{\frac{1}{\lambda+2}}-\lambda x_0^{\frac{1}{\lambda+2}}}\right],
\end{equation}
where $x_0$ is a solution of 
\begin{equation}\label{x0choice}
	x_0\, \left[\lambda-W\left(-x_0\right)\right]^{\lambda+2}=Y_\lambda.
\end{equation}

When the function $F_\lambda(x_0)=x_0\left[\lambda-W\left(-x_0\right)\right]^{\lambda+2}$ has a maximum, it is located at 
\begin{equation}\label{xsdef}
	\bar x_\lambda=a_\lambda\rme^{-a_\lambda},\quad {\rm with}\quad a_\lambda=\frac{1}{2}(\sqrt{9+4\lambda}+3).
\end{equation}
Notice that $(3-\sqrt{9+4\lambda})/2$ would also be a possible value for $a_\lambda$, but the choice~(\ref{xsdef}) ensures that $\bar x_\lambda$ is located in the domain of $x_0$. 
One finds the constraint
\begin{equation}\label{gmdef}
	Y_\lambda\leq \bar F_\lambda=F(\bar x_\lambda)=a_\lambda\rme^{-a_\lambda}(\lambda+a_\lambda)^{\lambda+2}.
\end{equation}

As for the exponential potential, (\ref{elneg}) vanishes for particular values of $Y_{\lambda}$, and thus critical heights appear. It can be checked that $\epsilon_\lambda$ is equal to zero when $Y_\lambda=2(\lambda+2)^{\lambda+2}/\rme^2$, which is lower than $\bar F_\lambda$ as demanded by (\ref{gmdef}). The critical heights are then given by 
\begin{equation}\label{gcneg}
	g_{\lambda;nl}=\left(\frac{\rme}{\lambda+2}\right)^{\lambda+2}\, N^2_{nl}.
\end{equation}
Again, the AFM gives no information about the best form for the number $N_{nl}$. A first trial is to use $N=N_\lambda$ but other choices are possible. The case $\lambda=0$ was discussed in the previous section. The case $\lambda=-1$ will be studied in the next section.

As a general result, (\ref{gcneg}) states that $g_{nl}\propto N^2_{nl}$ for any exponential potential. Interestingly, it has previously been observed that the critical heights of an arbitrary exponential potential can be very accurately fitted by a biquadratic form in $n$ and $l$ \cite{gsk1}. Our results can be seen as a demonstration of this empirical observation. However, the accuracy of the fitted form in \cite{gsk1} is a supplementary argument showing that the explicit form of $N_{nl}$ cannot be computed within the AFM: Only $N_{nl}=b n+l+c$ can be assumed, the coefficients $b$ and $c$ having to be determined by another procedure in order to improve the analytical formulae that we have obtained. 

Equations~(\ref{elneg}) and (\ref{x0choice}) formally define the analytical energy formula we are looking for. Finding a general solution of (\ref{x0choice}) is a complicated problem that cannot be solved analytically. Moreover, the Yukawa potential is, to our knowledge, the only potential of common use with $\lambda\neq0$. Before making a full treatment of the case $\lambda=-1$ in the next section, we can nevertheless make some comment about the limit where $\beta$ goes to zero ($g\to \infty$ and $Y_\lambda\to 0$). In this limit, $-\alpha\, r^\lambda \rme^{-\beta r}\approx -\alpha\, r^\lambda$, and we should recover (\ref{eq:eigenerplpot}) for $\lambda<0$ and $a=\alpha$. The nontrivial solution of (\ref{x0choice}) for $Y_\lambda=0$ and $\lambda<0$ is $x_0=-\lambda\, \rme^\lambda$. It can then be checked that 
\begin{equation}
\lim_{\beta\to 0} E_\lambda= 
\lim_{\beta\to 0}\frac{\beta^2}{2m}\epsilon_\lambda(-\lambda \rme^\lambda)=\frac{2+\lambda}{2 \lambda} (\alpha |\lambda|)^{2/(\lambda+2)}
\left ( \frac{N^2}{m} \right )^{\lambda/(\lambda+2)}.
\end{equation}
This is the expected result for the energy spectrum of the potential $-\alpha/r^{|\lambda|}$. It is important to notice that the trivial solution $x_0=0$ would lead to the unphysical limit $E_\lambda\to 0$. 

In the limit $\lambda\to 0$, equation~(\ref{x0choice}) reduces to
\begin{equation}
x_0 W(-x_0)^2=Y_0=(3Z)^3
\end{equation}
and its solution is given by
\begin{equation}
x_0=-3 W(-Z)\rme^{3 W(-Z)},
\end{equation}
in which $Z$ is defined by (\ref{defZ}) (paying no attention how the number $N$ is defined). Taking into account that $-Z \in[-1/\rme,0]$, a range in which $W(-Z) < 0$, and that (\ref{elneg}) reduces to
\begin{equation}
\epsilon_0=-g\, x_0\left[\sqrt{\frac{x_0}{Y_0}} - \frac{1}{2}\right],
\end{equation}
one recovers formula (\ref{epsnu0}), but with \emph{a priori} $N=N_0$. This is an illustration that the same functional form is obtained independently of the power-law chosen for the potential $P(x)$. Moreover, $g_{\lambda=0;nl}=g_{e;nl}$ and (\ref{gc1}) is also recovered.

Up to now, the branch to be used in these formulae is not yet determined. For $-2 < \lambda \leq -1$, the function $K(x)$ defined by (\ref{eqs}) is monotonic and its inverse is single valued; a careful inspection of its properties shows that only the $W_{-1}$ branch must be retained. For $\lambda > -1$, the situation is more complex and we did not find a simple prescription to fix unambiguously the type of branch. Fortunately the corresponding potentials are not of common use.
 
\section{Yukawa potential}\label{yuka}

The Yukawa potential corresponds to the case $\lambda=-1$. A look at (\ref{eqs}) in this case shows that one has $\nu_0\in[0,g]$ and equivalently $x_0\in[0,1/\rme]$. Since $x=I_{-1}(\nu)\in[0,\infty[$, $W_{-1}$ must be chosen: This branch of the Lambert function is defined on a finite domain, but can go to infinity as demanded.
In this case, eqs. (\ref{elneg}) and (\ref{x0choice}) read
\begin{equation}\label{eny2}
	\epsilon_y(\bar Y) = - g\frac{\rme\, x_0(\bar Y)^2 [x_0(\bar Y) - \bar Y]}{2\bar Y[x_0(\bar Y)+\bar Y]} ,
\end{equation}
\begin{equation}\label{eqelimy}
	-x_0(\bar Y)\, \left[1+W_{-1}(-x_0(\bar Y))\right]=\bar Y,
\end{equation}
where $\bar Y$ denotes $Y_{-1}=2N^2/(\rme\, g)$. 
Notice that, in principle, $N= N_{-1}$ in this last equality. But, as mentioned above, another choice $N_\eta$ with an arbitrary value of $\eta$ is \emph{a priori} possible.
The function $F_{-1}(x_0)=-x_0\, \left[1+W_{-1}(-x_0)\right]$ is such that $ F_{-1}(0)=F_{-1}(1/\rme)=0$, with a maximum in 
\begin{equation}
	\bar x=\frac{3+\sqrt 5}{2}\, \rme^{-\frac{3+\sqrt 5}{2}},\quad \bar F=F_{-1}(\bar x)=(2+\sqrt 5)\, \rme^{-\frac{3+\sqrt 5}{2}}.
\end{equation}
Let us note the relation $\bar F/\bar x$ = $\Phi$ = $(1+\sqrt{5})/2$, the golden ratio. Following (\ref{gcneg}), the critical heights read in this case 
\begin{equation}\label{ch1}
	g_{y;nl}=\rme\, N_{nl}^2.
\end{equation}

A previous formula giving analytical approximate energy levels for the Yukawa potential has been found in~\cite{hall} in the framework of the envelope theory. It is given in our notations by  
\begin{equation}
	E_{nl}^{{\rm up}}=\min_{x>0}\left(\frac{N_{-1}^2}{x^2}-g\frac{{\rm e}^{-x}}{x}\right).
\end{equation}
But, we show in~\ref{envyuk} that $E_{nl}^{{\rm up}}$ is exactly equal to $\epsilon_y(\bar Y)$ defined by (\ref{eny2}) and (\ref{eqelimy}). It means that the AFM and the envelope theory lead to the same approximate energy formula for the Yukawa potential, provided the quantity $N$ is restricted to $N_{-1}$. Nevertheless, a closed energy formula has not been proposed in~\cite{hall}; we achieve such a task in the following.  

\begin{figure}[t]
\begin{center}
\includegraphics*[width=9.0cm]{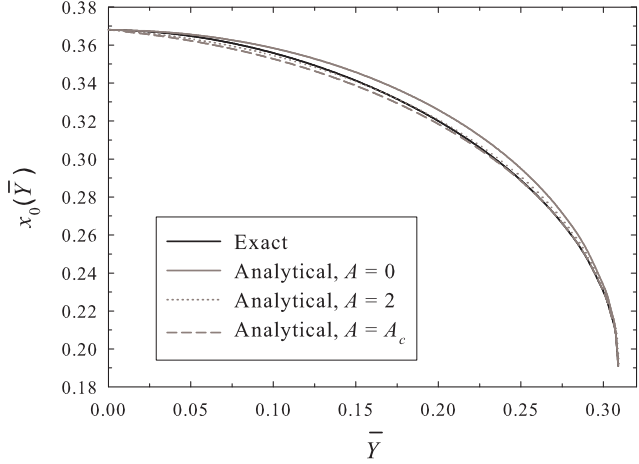}
\end{center}
\caption{Plot of the exact solution of (\ref{eqelimy}) (solid line), compared to the fitted form~(\ref{thfit}) with $A=0$ (full gray line), $A=2$ (dotted gray line), and $A=A_c\approx2.87$ (dashed gray line). The case $A=2$ is nearly indistinguishable from the exact solution.}
\label{fig2}
\end{figure}

In order to have an explicit energy formula, it is of interest to find an analytical approximation for $x_0(\bar Y)$, the solution of (\ref{eqelimy}). For a given value of $\bar Y$, two values of $x_0$ are allowed: The first one goes to zero when $\bar Y$ vanishes, and the second one tends to $1/\rme$ in this limit. We already mentioned that the physical solution should actually be such that $\lim_{ \bar Y\to 0}x_0(\bar Y)=1/\rme$ in order to find the correct Coulomb limit when $\beta\to 0$. From the study of the function $F_{-1}(x)$, it is possible to determine that $x_0(0)=1/\rme$ and $x_0(\bar F)=\bar x$. Moreover, in $\bar Y=0$ and $\bar Y=\bar F$, the function $x_0(\bar Y)$ has a horizontal and a vertical tangent respectively. So, the simplest form to parameterise $x_0(\bar Y)$ is a branch of ellipse.

As it can be seen in figure~\ref{fig2}, the analytical form 
\begin{equation}\label{thfit}
	x_0(\bar Y,A)=\bar x+\left(\frac{1}{\rme}-\bar x\right)\sqrt{1-\left(\frac{\bar Y}{\bar F}\right)^2+A\, \bar Y (\bar Y-\bar F)}
\end{equation}
reproduces rather well the numerical solution for the simplest choice $A=0$, but is particularly accurate for the optimal value $A=2$. When $A\ne 0$, the curve is not a branch of ellipse and the tangent in $\bar Y=0$ is not horizontal, but the global agreement is improved. Recalling that $g=2m\alpha/\beta$ for the Yukawa potential, the final energy spectrum reads 
\begin{equation}
\label{eyukA2}
\fl E_y=\frac{\beta^2}{2m}\epsilon_y(x_0(\bar Y,2))=- \alpha\,\beta\frac{\rme\, x_0(\bar Y,2)^2 [x_0(\bar Y,2) - \bar Y]}{2\bar Y[x_0(\bar Y,2)+\bar Y]},\quad {\rm with}\quad \bar Y= \frac{\beta N^2}{\rme\,m\alpha}.
\end{equation}
This formula can be recast in the following form
\begin{equation}
\label{eyukA2b}
E_y=E_{\rm Coul} \left[\rme^2 x_0(\bar Y,2)^2 \frac{x_0(\bar Y,2) - \bar Y}{x_0(\bar Y,2)+\bar Y}\right],
\end{equation}
where $E_{\rm Coul}$ is the energy for a pure Coulomb potential given by
\begin{equation}
\label{purecoul}
E_{\rm Coul} = -\frac{\alpha^2 m}{2 N^2}.
\end{equation}
As the term within brackets in (\ref{eyukA2b}) tends toward 1 when $\beta\to 0$, we can expect that $N=N_{-1}$ is a good choice for small values of $\beta$. This will be examined in section~\ref{impfore}.

Using (\ref{ch1}), one can write
\begin{equation}
\label{ybgc}
\bar Y=\frac{2}{\rme^2}\frac{g_{y;nl}}{g}.
\end{equation}
This makes clear the dependence on the critical heights $g_{y;nl}$ for formulae~(\ref{eyukA2}) and (\ref{eyukA2b}).
It is worth mentioning that $A=2$ comes from a fit of formula~(\ref{thfit}) on the numerical solution of~(\ref{eqelimy}). Another choice could have been done, that is to set $A=A_c$ such that 
\begin{equation}
x_0\left(\bar Y=\frac{2}{\rme^2},A_c\right)=\frac{2}{\rme^2}.	
\end{equation}
Then the final energy formula exactly vanishes at the critical heights (\ref{ch1}). One then obtains for $A_c$ a complicated analytical formula that we do not give explicitly here; it is enough for our purpose to know that $A_c\approx 2.87$.  

The results we obtained here in the most interesting case of the Yukawa potential can actually be generalised to potentials with $-2<\lambda\leq-1$. It can indeed be numerically checked that the solution of (\ref{x0choice}) in this range is quite well fitted by the analytical form
\begin{equation}
\fl	x_0( Y_\lambda,A_\lambda)=\bar x_\lambda+\left(|\lambda|\rme^{\lambda}-\bar x_\lambda\right)\sqrt{1-\left(\frac{Y_\lambda}{\bar F_\lambda}\right)^2+A_\lambda\, Y_\lambda (Y_\lambda-\bar F_\lambda)},
\end{equation}
with 
\begin{equation}\label{Adef}
A_\lambda=-(109+196\lambda+85\lambda^2).
\end{equation}
This function is such that $A_{-1}=2$.
Notice that the branch $W_{-1}$ has to be chosen, by continuity with the Yukawa case in particular. Once this solution is assumed, the energy spectrum is eventually given by $	\epsilon_\lambda(x_0(Y_\lambda,A_\lambda))$ [see (\ref{elneg}) with $W=W_{-1}$]. 

It has been shown in \cite{g1} that the critical heights of the Yukawa potential are given with a relative accuracy around 0.4\% by the following fitted form
\begin{equation}
\label{gnlemp}
	g^G_{nl}=2\left(\sqrt{Z_l}+\frac{n}{S_l}\right)^2,
\end{equation}
where 
\begin{eqnarray}
	&&Z_l=Z_0(1+\alpha l+\beta l^2),\quad S_l=S_0(1+\gamma l+\delta l^2), \nonumber \\
	&&Z_0=0.839908,\quad \alpha=2.7359,\quad \beta=1.6242, \nonumber \\
	&&S_0=1.1335,\quad \gamma=0.019102,\quad \delta=-0.001684. 
\end{eqnarray}
Due to the smallness of $\gamma$ and $\delta$, we can replace $S_l$ by $S_0$ in (\ref{gnlemp}).
It clearly appears that $g_{nl}\propto(S_0\sqrt{Z_0\beta}\, l+n)^2$ asymptotically. This quadratic behaviour is predicted by our analytical results. Introducing a constant which allows to obtain the exact result for $n=l=0$ and considering the asymptotic expansion for large values of $l$, 
(\ref{gnlemp}) is approximately given by
\begin{equation}
\label{gnlemp2}
	g^G_{nl}\approx \left( 1.248\, n+1.652\, l+1.296\right)^2.
\end{equation}

It could be interesting to improve the values of critical heights for the Yukawa potential as it was made for the pure exponential potential. Unfortunately, no analytical formula for $g_{y;n0}$ is available in this case. Nevertheless, we can assume that the formula~(\ref{ch1}) is a good starting point and that $N_{nl}= b\, n+l+c$ with no constraint on parameters $b$ and $c$ (contrary to what is expected from (\ref{eq:grandNl})). If the two exact values $g^*_{y;00}$ and $g^*_{y;10}$ are known, the choice $b=\sqrt{g^*_{y;10}/\rme}-\sqrt{g^*_{y;00}/\rme}$ and $c=\sqrt{g^*_{y;00}/\rme}$ insures that (\ref{ch1}) gives the correct critical heights for $l=0$ and $n=0$ and 1. Instead of using exact numerically computed values, we can use analytical approximate evaluations of these quantities (see \ref{gyuk}). We can then try a new relation to compute the critical heights 
\begin{equation}\label{ch1b}
	g_{y;nl}=\left( \left[\sqrt{g_{y;10}}- \sqrt{g_{y;00}}\right] n+ \sqrt{\rme}\, l+ \sqrt{g_{y;00}}\right)^2.
\end{equation}
With the exact critical heights, we have
\begin{equation}\label{ch1c}
	g_{y;nl}\approx \left( 1.243\, n+ 1.649\, l+1.296 \right)^2,
\end{equation}
while with the use of approximations~(\ref{gyuk2}), we have
\begin{equation}\label{ch1d}
	g_{y;nl}\approx \left( 1.291\, n+ 1.649\, l+ 1.296\right)^2.
\end{equation}
These formulae are in good agreement with (\ref{gnlemp2}) and with exact results: For $n\in[0,4]$ and $l\in[0,4]$, the minimal, maximal and mean relative errors are respectively 0.005\%, 4.5\% and 1.2\% with (\ref{ch1c}) and respectively 0.04\%, 5.8\% and 3.4\% with (\ref{ch1d}). It is not possible to obtain so good agreement by choosing $b$ and $c$ values coming from (\ref{bcdef}). Let us note that the minimal, maximal and mean relative errors are respectively 0.01\%, 0.9\% and 0.3\% with formula~(\ref{gnlemp}). The very good quality of these results is due to the use of a complicated formula fitted on exact results. Our formulae are simpler and the general behavior is predicted by the AFM. Some exact $g^*_{y;nl}$ values are given in table~\ref{tab1}; they are computed numerically by solving the Schr\"{o}dinger equation with a vanishing energy

\begin{table}[ht]
\caption{\label{tab1}Some exact critical heights $g^*_{y;nl}$ for the Yukawa potential $-g\, \rme^{-x}/x$.}
\begin{indented}
\item[]\begin{tabular}{@{}ccccc}
\br 
$n$ & $l=0$ & 1 & 2 & 3\\
\mr
0 & 1.680&  9.082 & 21.895 & 40.136\\
1 & 6.447 & 17.745 & 34.420 & 56.514\\
2 & 14.342&  29.461 & 49.970 & 75.899\\
3 & 25.372 & 44.261 & 68.572 & 98.318\\
\br
\end{tabular}
\end{indented}
\end{table}

In the very same way that we improved the critical heights of the pure exponential potential, it is possible to improve the corresponding Yukawa quantities with the same kind of prescription. Here we propose the expression
\begin{equation}
\label{ch1e}
g_{y;nl}\approx \left( 1.247\, n+ 1.680\, l-0.054 \sqrt{n\,l}+ 1.296\right)^2.
\end{equation}
In this case, the minimal, maximal and mean relative errors are respectively 0.002\%, 2.4\% and 0.6\%. We obtain a quality of the same order than the one obtained with formula~(\ref{gnlemp}), but with a much simpler expression.

\section{Improved energy formulae}\label{impfore}

In the previous sections, approximate analytical forms for solutions of a pure 
exponential or a Yukawa potentials were found. The formulae depend on the 
quantum numbers $n$ and $l$ through a factor $N$. This number could be taken 
as $N_\lambda$ without information about the optimal value for $\lambda$. If we
look for instance at the Yukawa potential, it is clear that
it reduces to a pure Coulomb interaction when $\beta=0$. In this case, the choice
$N=N_{-1}$ gives the exact result in this limit. 

The dimensionless Hamiltonians considered above depend on a parameter $g$. The
variation of the eigenvalues being smooth for the variation of $g$, we can
assume that the number $N$, giving the optimal values for all the eigenvalues
of a Hamiltonian, is also a smooth function of $g$. From considerations above, the
functional form 
\begin{equation}
\label{Nbeta}
N(g)= b(g)\, n + l + c(g) 
\end{equation}
seems reasonable. As we cannot predict the correct behaviour, it is 
necessary to focus our attention on numerical solutions. 
 
Very accurate eigenvalues $\epsilon_{\textrm{num}}(g;n,\ell)$ for Hamiltonians
defined above can be obtained numerically with the Lagrange mesh method \cite{lag}.
It is very accurate and easy to implement. In order to find the best possible
values for coefficients $d(g)$ ($d$ stands for $b$ or $c$), we will use the
following measure
\begin{equation}
\label{chi2}
\chi(g)=\sum_{\{n,l\}} \left( \epsilon_{\textrm{num}}
(g;n,\ell) - \epsilon_{\textrm{app}}(g;n,\ell) \right)^2,
\end{equation}
where $\epsilon_{\textrm{app}}(g;n,\ell)$ are values obtained from our
approximate formulae.
Other choices are possible but we find this one very convenient. The summation in
(\ref{chi2}) runs \emph{a priori} on all possible pairs $\{n,l\}$ for the bound states
allowed by the value of $g$. But for some special pairs $\{n,l\}$, the value 
$\epsilon_{\textrm{app}}(g;n,\ell)$ can be a complex number 
(condition~(\ref{x0choice}) is not fulfilled) or a positive number 
(quality of approximation is poor). This drawback can happen for the highest levels. In this case,
the pair $\{n,l\}$ is not taken into account in (\ref{chi2}).

The analytical form $\epsilon_{\textrm{app}}$ depends on $N(g)$ which depends
on coefficients $d$. 
For each value of $g$, optimal values for the $d$ coefficients, $d_{\textrm{min}}
(g)$, can be determined by minimizing $\chi(g)$. Then, with a set
\{$d_{\textrm{min}}(g)$\} for a given set \{$g$\}, a functional form
$d_{\textrm{fit}}(g)$ can be fitted with the following measure ($\chi'$ is also a chi-square
fit, but the prime indicates that the function $\chi'$ is indeed different from $\chi$ in
(\ref{chi2}))
\begin{equation}
\label{chi2bis}
\chi'(d)=\sum_{\{g\}} \left( d_{\textrm{min}}
(g) - d_{\textrm{fit}}(g) \right)^2.
\end{equation}
Again, other choices are possible but we find this one very convenient. We will now
try to determine the best form of coefficients $d(g)$ for the potentials
studied above. 

\subsection{Pure exponential potential}

\begin{figure}[ht]
\begin{center}
\includegraphics*[width=6.4cm]{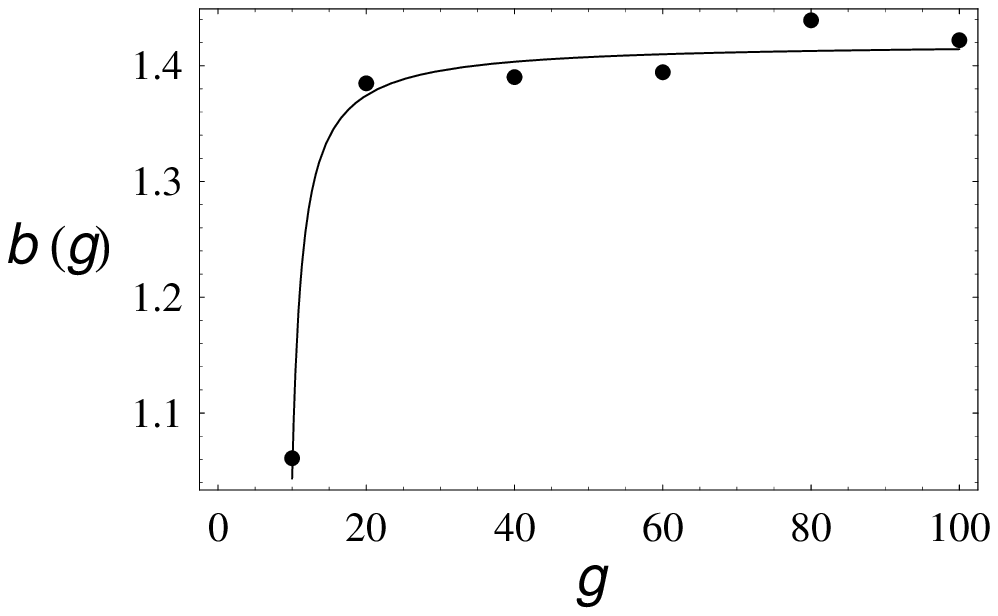}
\includegraphics*[width=6.4cm]{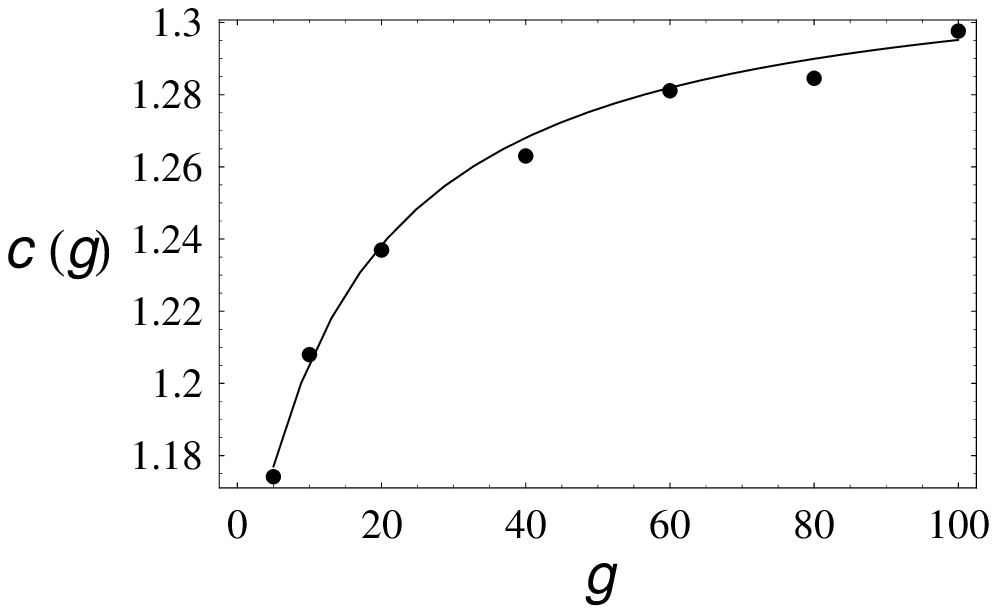}
\caption{\label{fig:exp} Best values of the coefficients $b(g)$ and $c(g)$
to parameterise the eigenvalues~(\ref{epsnu0}) for a pure exponential potential: 
numerical fit with (\ref{chi2}) (dots); functions (\ref{eq:bcexp}) (solid line).} 
\end{center}
\end{figure}

Approximate dimensionless eigenvalues for 
the pure exponential potential are given by (\ref{epsnu0}).
By minimising our measure $\chi(g)$, we found the optimal values of $b(g)$ 
and $c(g)$ for several finite values of $g$. The results are plotted with dots on 
figure~\ref{fig:exp}. The small irregularities in the data are due to the fact 
that the number of bound states is obviously not a smooth function of $g$.

We tried to fit the numerical points with various functions and found that the best
result is obtained for sections of hyperbola, for both $b(g)$ and $c(g)$. The
parameterisation retained is 
\begin{equation}
\label{eq:bcexp}
b(g)= \frac{1.42\, g-12.76}{g-8.62}, \quad 
c(g)= \frac{1.32\, g+16.88}{g+14.95}.
\end{equation}
The agreement with numerical points is good as it can be seen on figure~\ref{fig:exp}.
For large value of $g$, a constant value of the coefficient $b$ is a good approximation. 
This is clearly not the case for the coefficient $c$. The quality of this parameterisation
for both $b(g)$ and $c(g)$ in (\ref{epsnu0}) can be appraised in tables~\ref{tab:exp1}
and \ref{tab:exp2}. 
Not all bound states can be found. As in (\ref{chi2}), only real strictly negative eigenvalues are 
retained. The mean
relative error is about 20\% but the accuracy is much better for the lowest eigenvalues. 

\begin{table}[htb]
\caption{\label{tab:exp1} Quality of various approximations for a pure exponential potential.
$R$: (number of negative eigenvalues found)/(total number of bound states);
$\Delta_{\textrm{min./max.}}$: minimal/maximal relative error (\%) for 
negative eigenvalues found; $\Delta_{\textrm{mean}}$: mean relative error (\%)  
computed with all negative eigenvalues found.
First line: formula~(\ref{epsnu0}) with parameterisation~(\ref{eq:bcexp});
Second line: formula~(\ref{epsnu0}) with parameterisation~(\ref{eq:abcdexp}).}
\begin{indented}
\item[]\begin{tabular}{@{}cccccccc}
\br
$g$ & 5 & 10 & 20 & 40 & 60 & 80 & 100 \\
\mr
$R$ & 1/1 & 3/3 & 4/6 & 7/10 & 12/15 & 15/20 & 19/26 \\
    & 1/1 & 3/3 & 4/6 & 8/10 & 12/15 & 16/20 & 21/26 \\
$\Delta_{\textrm{min.}}$ & 0.8 & 4.1 & 1.7 & 1.6 & 0.4 & 0.5 & 0.3 \\
                         & 2.0 & 1.8 & 0.05 & 0.3 & 0.07 & 0.09 & 0.2 \\
$\Delta_{\textrm{max.}}$ & 0.8 & 54.6 & 79.7 & 35.0 & 98.5 & 91.2 & 81.7 \\
                         & 2.0 & 33.4 & 6.2 & 31.7 & 80.1 & 69.1 & 97.4 \\
$\Delta_{\textrm{mean}}$ & 0.8 & 31.6 & 22.6 & 11.8 & 23.6 & 18.3 & 19.1 \\
                         & 2.0 & 14.0 & 2.4 & 8.9 & 11.8 & 11.2 & 17.2 \\
\br
\end{tabular}
\end{indented}
\end{table}

By using the form~(\ref{Nbeta}), we restrict the number of functions to fit and we keep the same level of numerical complexity as the one used in \cite{af2}. Since relation~(\ref{gc1c}) works very well to predict the critical heights, we used the same form to fit the eigenvalues. Performing again similar calculations as above with (\ref{chi2}) and (\ref{chi2bis}), we found that a good parameterisation is given by
\begin{eqnarray}
\label{eq:abcdexp}
N(g)&=& \frac{1.44\, g-11.17}{g-6.86} n +\frac{0.95\, g-1.36}{g-0.33} l
+\frac{1.43\, g+23.09}{g+20.60} \nonumber \\ 
&&+ (6.69~10^{-6}g^2-0.00019\, g-0.126) \sqrt{n\, l}.
\end{eqnarray}
Its quality can be appraised in tables~\ref{tab:exp1} and \ref{tab:exp2}. The eigenvalues are generally improved and the number of bound states found is higher. Note that the influence of the term $\sqrt{n\, l}$ is small. The neglect of this term does not spoil very much the results.

When $\beta\to 0$, that is $g\to\infty$, the pure exponential potential reduces at first order to a constant and at the next order to a linear potential. Relations~(\ref{eq:bcexp}) show that $\lim_{g\to\infty}b(g)=1.42$ and that $\lim_{g\to\infty}c(g)=1.32$. These numbers are different from the values $b\approx 1.79$ and $c\approx 1.38$ expected for a linear potential, and from $b\approx 1.48$ and $c\approx 1.21$ expected for constant potential (see~(\ref{bcdef})). We will see that the situation will be clearer for the Yukawa potential. 

\begin{table}[htb]
\caption{\label{tab:exp2} Eigenvalues for a pure exponential potential with $g=40$
as a function of $\{ l,n\}$ sets. First line: exact value; Second line:
formula~(\ref{epsnu0}) with parameterisation~(\ref{eq:bcexp});
Third line: formula~(\ref{epsnu0}) with parameterisation~(\ref{eq:abcdexp}). 
A * indicates a non real or a non negative value.}
\begin{indented}
\item[]\begin{tabular}{@{}ccccc}
\br
$l$ & $n=0$ & 1 & 2 & 3 \\
\mr
0 & $-17.53$ & $-6.88$ & $-1.87$ & $-0.077$ \\
  & $-17.84$ & $-7.58$ & $-1.59$ & * \\
  & $-17.30$ & $-7.28$ & $-1.43$ & * \\
1 & $-10.14$ & $-3.35$ & $-0.42$ & - \\
  & $-9.98$ & $-2.99$ & * & - \\
  & $-10.10$ & $-3.55$ & * & - \\
2 & $-5.03$ & $-0.93$ & - & - \\
  & $-4.63$ & * & - & - \\
  & $-5.06$ & $-0.64$ & - & - \\
3 & $-1.55$ & - & - & - \\
  & $-1.01$ &- & - & - \\
  & $-1.52$ &- & - & - \\
\br
\end{tabular}
\end{indented}
\end{table}

\subsection{Yukawa potential}

\begin{figure}[ht]
\begin{center}
\includegraphics*[width=6.4cm]{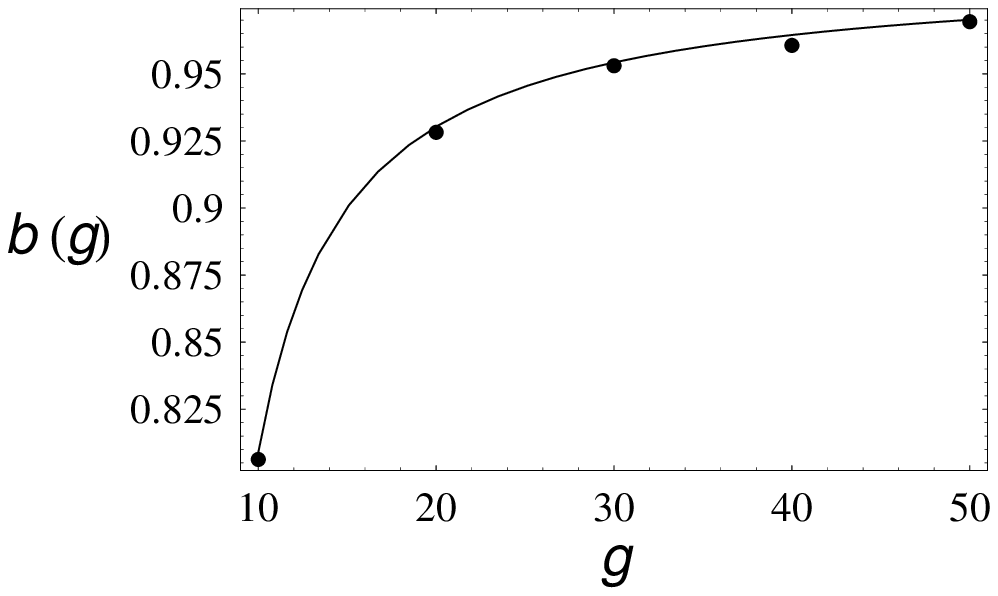}
\includegraphics*[width=6.4cm]{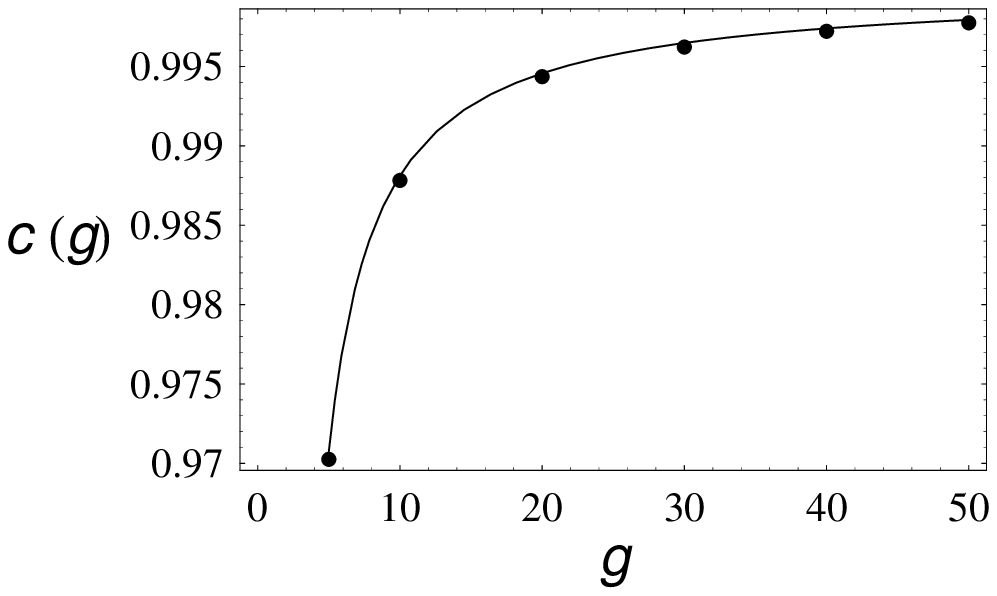}
\caption{\label{fig:yuk} Best values of the coefficients $b(g)$ and $c(g)$
to parameterise the eigenvalues~(\ref{eny2}) with (\ref{thfit}) for a Yukawa potential: 
numerical fit with (\ref{chi2}) (dots); functions (\ref{eq:bcyuk}) (solid line).} 
\end{center}
\end{figure}

Approximate dimensionless eigenvalues for 
the Yukawa potential are given by (\ref{eny2}) supplemented by (\ref{thfit}).
In all results presented, we chose $A=2$.
By using a similar procedure as those described for the pure exponential potential,
we found that the best result in fitting the numerical points is obtained 
for sections of hyperbola, for both $b(g)$ and $c(g)$. The
parameterisation retained is 
\begin{equation}
\label{eq:bcyuk}
b(g)= \frac{0.99\, g-5.92}{g-5.08}, \quad 
c(g)= \frac{1.00\, g-1.68}{g-1.58}.
\end{equation}
The agreement with numerical points is good as it can be seen on figure~\ref{fig:yuk}.
As expected, $N(g)\to N_{-1}$ when $g\to \infty$, that is to say when $\beta\to 0$, since in 
this limit the Yukawa potential tends toward a pure Coulomb interaction. So, in the following,
we will also consider the choice $N=N_{-1}$, that is $b=c=1$. Again a more complex parameterisation 
\begin{eqnarray}
\label{eq:abcdyuk}
N(g)&=& \frac{0.99\, g-7.16}{g-6.64} n +\frac{1.00\, g+2.51}{g+3.16} l
+\frac{1.00\, g-1.89}{g-1.79} \nonumber \\ 
&&+ (-0.000233\,g^2+0.0202\, g-0.480) \sqrt{n\, l}.
\end{eqnarray}
is also studied.

An accurate energy formula has been found in \cite{g1} from a fit of the numerically computed energy levels of the Yukawa potential. In our notations, it reads 
\begin{equation}\label{enap}
\epsilon_{nl}=-\frac{g}{4N^2_{nl}}(g-g^G_{nl}) \frac{g-2 A' (N_{nl}+\sigma)^2+2 B' N^2_{nl}}{g-g^G_{nl}+2 B' N^2_{nl}},
\end{equation}
where $A'=1.9875$, $B'=1.2464$ and $\sigma=0.003951$, where $N_{nl}=n+l+1$ as in the Coulomb case, and where $g^G_{nl}$ is defined by (\ref{gnlemp}). This formula is rather different from ours, given by (\ref{eny2}). However, they coincide at the limit $g\to\infty$. 

\begin{table}[htb]
\caption{\label{tab:yuk1} Quality of various approximations for Yukawa potential.
$R$: (number of negative eigenvalues found)/(total number of bound states);
$\Delta_{\textrm{min./max.}}$: minimal/maximal relative error (\%) for 
negative eigenvalues found; $\Delta_{\textrm{mean}}$: mean relative error (\%)  
computed with all negative eigenvalues found.
First line: formula~(\ref{eny2}) and (\ref{thfit}) with 
parameterisation~(\ref{eq:bcyuk}); Second line: formula~(\ref{eny2}) and (\ref{thfit}) with 
$N=N_{-1}$; Third line: formula~(\ref{eny2}) and (\ref{thfit}) with 
parameterisation~(\ref{eq:abcdyuk});
Fourth line: formula~(\ref{enap}). The abnormally large values marked by $\dagger$
are due to the existence of a nearly zero eigenvalue which is badly reproduced.
}
\begin{indented}
\item[]\begin{tabular}{@{}cccccccc}
\br
$g$ & 5 & 10 & 20 & 30 & 40 & 50 \\
\mr
$R$ & 1/1 & 2/3 & 3/5 & 6/8 & 6/10 & 10/13 \\
    & 1/1 & 1/3 & 3/5 & 6/8 & 6/10 & 10/13 \\
    & 1/1 & 2/3 & 3/5 & 6/8 & 6/10 & 10/13 \\
    & 1/1 & 3/3 & 5/5 & 8/8 & 10/10 & 12/13 \\
$\Delta_{\textrm{min.}}$ & 0.2 & 0.07 & 0.0006 & 0.02 & 0.02 & 0.02 \\
                         & 13.5 & 3.5 & 1.3 & 0.8 & 0.6 & 0.5 \\
                         & 0.7 & 0.2 & 0.01 & 0.01 & 0.02 & 0.02 \\
                         & 0.4 & 0.1 & 0.06 & 0.04 & 0.03 & 0.02 \\
$\Delta_{\textrm{max.}}$ & 0.2 & 4.3 & 6.6 & 27.8 & 9.4 & 42.4 \\
                         & 13.5 & 3.5 & 14.1 & 45.6 & 19.2 & 57.3 \\
                         & 0.7 & 8.6 & 0.7 & 30.0 & 9.8 & 44.1 \\
                         & 0.4 & 10.9 & 14.0 & 34.9 & 575.2$^{\dagger}$ & 24.6 \\
$\Delta_{\textrm{mean}}$ & 0.2 & 2.2 & 3.2 & 13.0 & 4.6 & 16.2 \\
                         & 13.5 & 3.5 & 7.7 & 19.6 & 8.2 & 21.2 \\
                         & 0.7 & 4.4 & 0.3 & 6.8 & 2.1 & 11.1 \\
                         & 0.4 & 6.4 & 3.3 & 9.8 & 59.5$^{\dagger}$ & 3.9 \\
\br
\end{tabular}
\end{indented}
\end{table}

\begin{table}[htb]
\caption{\label{tab:yuk2} Eigenvalues for a Yukawa potential with $g=30$
as a function of $\{ l,n\}$ sets. First line: exact value. Second line:
formula~(\ref{eny2}) and (\ref{thfit}) with 
parameterisation~(\ref{eq:bcyuk}). 
Third line: formula~(\ref{eny2}) and (\ref{thfit}) with $N=N_{-1}$;
Fourth line: formula~(\ref{eny2}) and (\ref{thfit}) with 
parameterisation~(\ref{eq:abcdyuk});
Fifth line: formula~(\ref{enap}).
A * indicates a non real or a non negative value.}
\begin{indented}
\item[]\begin{tabular}{@{}ccccc}
\br
$l$ & $n=0$ & 1 & 2 & 3 \\
\mr
0 & $-196.44$ & $-31.51$ & $-5.47$ & $-0.22$ \\
  & $-196.41$ & $-32.31$ & $-4.16$ & * \\
  & $-194.82$ & $-29.66$ & $-2.97$ & * \\
  & $-196.42$ & $-31.74$ & $-3.89$ & * \\
  & $-196.36$ & $-31.62$ & $-5.67$ & $-0.30$ \\
1 & $-30.74$ & $-4.94$ & $-0.029$ & - \\
  & $-29.84$ & $-3.57$ & * & - \\
  & $-29.66$ & $-2.97$ & * & - \\
  & $-30.89$ & $-4.82$ & * & - \\
  & $-30.52$ & $-4.85$ & $-0.019$ & - \\
2 & $-3.81$ & - & - & - \\
  & $-3.02$ & - & - & - \\
  & $-2.97$ & - & - & - \\
  & $-3.49$ & - & - & - \\
  & $-3.70$ & - & - & - \\ 
\br
\end{tabular}
\end{indented}
\end{table}

The quality of the various approximations can be appraised in tables~\ref{tab:yuk1}
and \ref{tab:yuk2}. Not all bound states can be found, even if formula (\ref{enap}) 
gives generally the correct number of bound states. As in (\ref{chi2}), 
only real strictly negative eigenvalues are retained. 
Formula~(\ref{eny2}) with approximation (\ref{thfit}) and parameterisation~(\ref{eq:bcyuk})
gives quite good results. The mean relative error is about 10\% but the accuracy is much better for the lowest eigenvalues. The parameterisation~(\ref{eq:abcdyuk}) gives generally better results.
The quality of the fit for formula~(\ref{enap}) and our approach is in average quite comparable. But formula~(\ref{enap}) is empirical. Conversely, (\ref{eny2}) is obtained from an explicit analytical resolution of the Schr\"{o}dinger equation. Our results are not so good by choosing the parameterisation $N=N_{-1}$ but the quality is still reasonable.  

\section{Summary of the results}\label{conclu}

The AFM was proposed in \cite{af,af2} as a tool to compute approximate 
analytical solutions of the Schr\"{o}dinger equation. The basic idea underlying this 
method is to replace an arbitrary potential $V(r)$, for which no 
analytical spectrum is known, by an expression of the type 
$\nu\, P(r)+g(\nu)$, $P(r)$ being a potential for which analytical 
eigenenergies can be found, $\nu$ the auxiliary field and $g(\nu)$ a 
well-defined function of this
extra parameter. This auxiliary field is such 
that its elimination as an operator leads to the original Hamiltonian. 
If $\nu$ is seen as a number however, analytical approximate solutions
can be found in favourable cases, and the auxiliary field is eventually 
eliminated by a minimisation on the eigenenergies. It is shown in \cite{af,af2}
that this method is a kind of mean field approximation.

In the present work, we have searched for solutions of the Schr\"{o}dinger equation with exponential potential of the form $-\alpha\, r^\lambda \rme^{-\beta\, r}$ using the AFM. General analytical results have been obtained for the eigenvalues of this type of potentials. It is proved that the critical heights for these interactions are proportional, with a good approximation, to $(b\, n+l+c)^2$ where $b$ and $c$ are constant parameters. 

We have focused our attention on pure exponential and Yukawa interactions. Approximate analytical formulae have been found for these two particular cases for arbitrary values of $n$ and $l$. Some results were previously obtained for the Yukawa potential \cite{hall,gara91} but, up to our knowledge, it is the first time that a closed energy formula is computed from the Schr\"{o}dinger equation. The majority of eigenvalues can be computed. The accuracy is quite good, mainly for lowest states. Moreover, very simple analytical approximate formulae have been obtained for the critical heights of these two potentials. 

As an outlook, we mention that more complicated problems could be studied with the AFM. In particular, we plan to apply this method to semirelativistic Hamiltonians. Such a work is in progress. Moreover, we have shown that the energy spectrum we get with the AFM for the Yukawa potential is exactly equal to the one which is obtained with the envelope theory. This suggests the existence of a connection between both methods, that we will discuss in a subsequent paper.   

\ack

CS and FB thank Francis Michel for a helpful discussion. They also thank the F.R.S.-FNRS for financial support.

\begin{appendix}
\section{The Lambert function}\label{lambprop}
\begin{figure}[t]
\begin{center}
\includegraphics*[width=8cm]{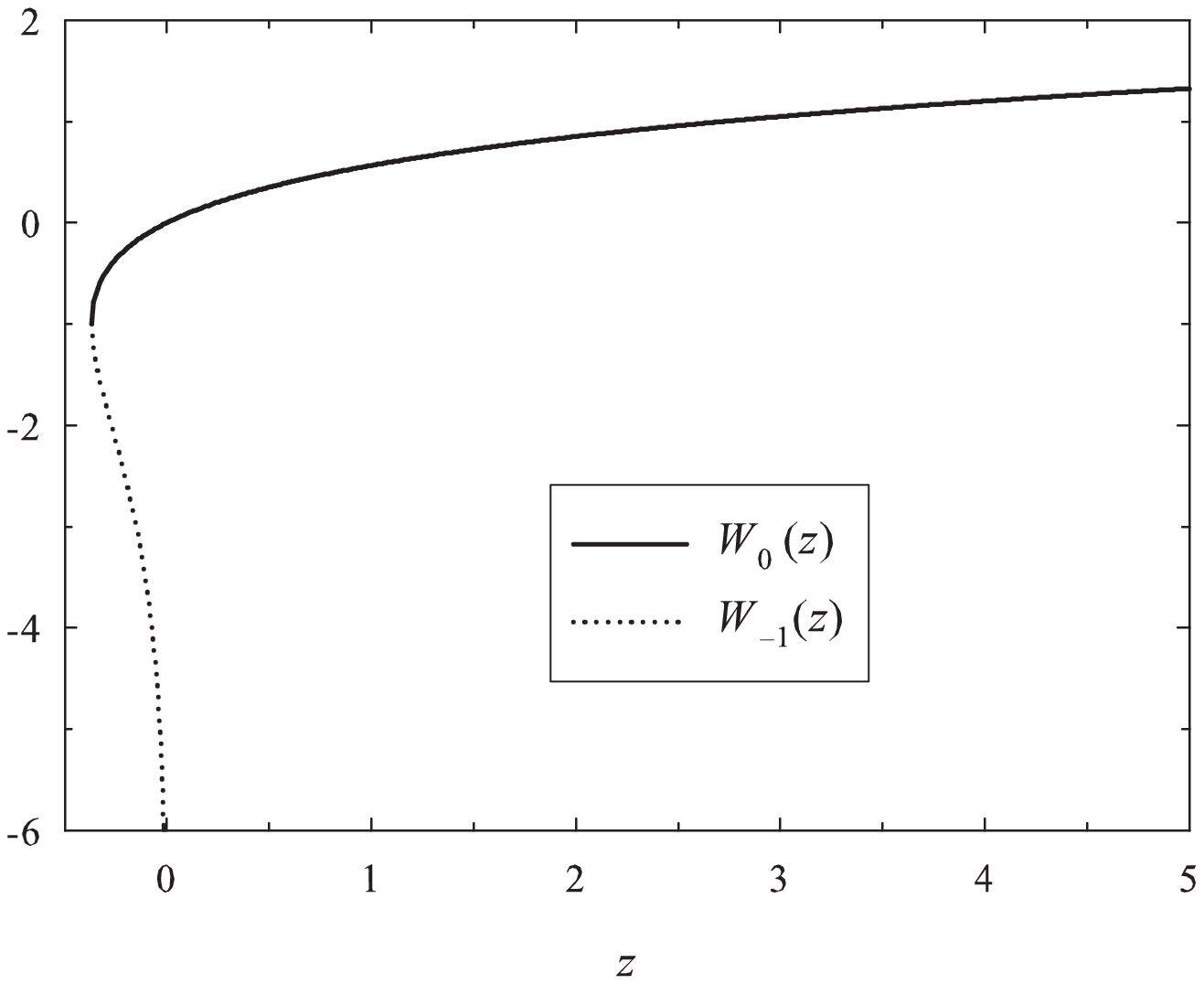}
\end{center}
\caption{Plot of the two branches of the Lambert function, namely $W_0(z)$ (solid line) and $W_{-1}(z)$ (dotted line). }
\label{fig1} 
\end{figure}

Let us briefly recall some points concerning the Lambert function (also called Omega function or product-log), that we will denote $W(z)$. First of all, $W(z)$ is defined as the inverse function of $z\, \rme^z$. Consequently, it has the following properties:
\begin{equation}
\label{defLamb}
	W(z)\, \rme^{W(z)}=W\left(z\, \rme^z\right)=z,
\end{equation}
\begin{equation}
\label{derivLamb}
	\partial_z\, W(z)=\frac{W(z)}{z\left[1+W(z)\right]}.
\end{equation}
But, it is readily observed that the inverse function of $z\, \rme^z$ is multivalued. Two branches of the Lambert function thus exist, respectively denoted as $W_0(z)$, defined for $z\geq-1/\rme$, and $W_{-1}(z)$, defined for $-1/\rme\leq z\leq0$ \footnote{Notice that the two branches $W_0(x)$ and $W_{-1}(x)$ of the Lambert function $W(x)$ are known by the software \emph{Mathematica} package as {\ttfamily ProductLog[0,x]} and {\ttfamily ProductLog[-1,x]} respectively.}. They are plotted in figure~\ref{fig1}. Obviously, both branches of the Lambert function share the same properties (\ref{defLamb}) and (\ref{derivLamb}) as $W(z)$; they meet in $W_{-1}(-1/\rme)=W_0(-1/\rme)=-1$. It can moreover be checked that $W_0(|x|\ll 1)\approx x$, $\lim_{x\to 0}W_{-1}(x)=-\infty$, and $\lim_{x\to \infty}W_{0}(x)=\infty$.

For our purpose, it is worth mentioning that the equation 
\begin{equation}\label{resol}
	(a\, z+b)^n\, \rme^{-z}=\theta
\end{equation}
is analytically solvable with the Lambert function. One has 
\begin{eqnarray}\label{solulamb}
	z&=&-\frac{b}{a}-n\, W\left[-\frac{1}{a\, n}(\rme^{-b/a}\theta)^{1/n}\right],
\end{eqnarray}
where either $W_0$ or $W_{-1}$ has to be chosen following the range of $\theta$ and $z$ that the function $z(\theta)$ has to cover. 

\section{Yukawa potential and envelope theory}\label{envyuk}
The envelope theory is a powerful method allowing to get upper and lower bounds on the energy levels of a given potential~\cite{env0}. The case of the Yukawa potential has been studied in~\cite{hall}. In particular, it is shown that an upper bound of the energy levels is given in our notations by
\begin{equation}
	E_{nl}^{{\rm up}}=\min_{x>0}f(x) \quad {\rm with}\quad f(x)=\frac{N_{-1}^2}{x^2}-g\frac{{\rm e}^{-x}}{x}.
\end{equation}
Equivalently, one can write the energy formula as
\begin{equation}
	E_{nl}^{{\rm up}}=f(\theta) \quad {\rm with}\quad	f'(\theta)=0.
\end{equation}
Explicitly, these relations can be rewritten as
\begin{equation}
	E_{nl}^{{\rm up}}=-\frac{N_{-1}^2}{\theta^2}\left(\frac{1-\theta}{1+\theta}\right),
\end{equation}
\begin{equation}
	\theta(\theta+1)\, {\rm e}^{-(\theta+1)}=\bar Y.
\end{equation}

These last two equations are completely equivalent to (\ref{eny2}) and (\ref{eqelimy}), giving the energy spectrum of the Yukawa potential obtained with the AFM, provided that the following redefinition is made
\begin{equation}
	x_0=(\theta+1)\, {\rm e}^{-(\theta+1)},
\end{equation}
and that the quantity $N_{-1}$ is replaced by $N$.
 
\section{Analytical estimation of critical heights $g_{n0}$ for the Yukawa potential}
\label{gyuk}

In \cite{hult51}, it is shown that, for a fixed value of the parameter $a$, solutions of the equation 
\begin{equation}
\label{gyuk1}
\frac{d^2\phi}{dx^2}+\left( a+b \frac{\rme^{-x}}{x}\right)\phi = 0,
\end{equation}
with the boundary conditions $\phi(0)=\phi(\infty)=0$, can be accurately obtained by a variational method. The eigenvalues $b$ are given by the extremum of the ratio $J/N$, where the quantities $J$ and $N$ are respectively given by equations (18) and (19) in \cite{hult51} (note a misprint: parameter $b$ must not be present in (19)). This ratio $J/N$ depends on $a$ and on $n+1$ linear variational parameters $h_\nu$. The critical heights $g_{n0}$ we search for are given by $\lim_{a\to 0} b$. 

For $n=0$, the only solution, which is independent of $h_0$ (normalisation condition), is given by $g_{00}=1/\ln(16/9)\approx 1.738$. This value if not too far from the exact one (see table~\ref{tab1}). For $n=1$, two extrema for $J/N$ exist and depend on $h_1/h_0$. They can be obtained analytically, but their expressions are very complicated and unusable in practice. Nevertheless, the two optimal values for $h_1/h_0$ are very close to $-2$ and $1/2$ respectively. So we can set
\begin{eqnarray}
\label{gyuk2}
g_{00}&\approx \left. \displaystyle\frac{J}{N}\right|_{h_1/h_0=1/2} &= \frac{17}{6 \ln(27/5)} \approx 1.680, \nonumber \\
g_{10}&\approx \left. \displaystyle\frac{J}{N}\right|_{h_1/h_0=-2}  &= \frac{1}{60 \ln(2)-26 \ln(3)-8 \ln(5)} \approx 6.693 .
\end{eqnarray}
These two values are good approximations of the exact ones.

\end{appendix}


\section*{References}

\begin{thebibliography}{99}
\bibitem{flu} Fl\"{u}gge S 1999 \textit{Practical Quantum Mechanics} (Berlin: Springer) and references therein
\bibitem{af} Silvestre-Brac B, Semay C and Buisseret F 2008 \textit{J. Phys. A: Math. Theor.} \textbf{41} 275301 (\textit{Preprint} arXiv:0802.3601)
\bibitem{af2} Silvestre-Brac B, Semay C and Buisseret F 2008 \textit{J. Phys. A: Math. Theor.} \textbf{41} 425301  (\textit{Preprint} arXiv:0806.2020)
\bibitem{af1} Dirac P A M 1964 \textit{Lectures on Quantum Mechanics} (New York: Belter Graduate School of Sciences, Yeshiva University)
\item[] Brink L, Di Vecchia P and Howe P S 1977 \textit{Nucl. Phys.} B {\bf 118} 76
\bibitem{lucha} Lucha W, Sch\"{o}berl F and Gromes D 1991 \textit{Phys. Rept.} \textbf{200} 1
\bibitem{gonz} Gonzalez P, Valcarce A, Vijande J and Garcilazo H 2005 \textit{Int. J. Mod. Phys.}  A {\bf 20} 1842 (\textit{Preprint} hep-ph/0409202) and references therein .
\bibitem{brau1} Brau F and Calogero F 2003 \textit{J. Math. Phys.} \textbf{44} 1554
\item[] Brau F and Calogero F 2003 \textit{J. Phys. A: Math. Gen.} \textbf{36} 12021
\item[] Brau F 2003 \textit{J. Phys. A} \textbf{36} 9907 (\textit{Preprint} math-ph/0401023)
\item[] Brau F and Lassaut M 2004 \textit{J. Phys. A: Math. Gen.} \textbf{37} 11243 (\textit{Preprint} math-ph/0411005)
\item[] Brau F 2005 \textit{J. Math. Phys.} \textbf{46} 032305 (\textit{Preprint} math-ph/0412042)
\bibitem{Abra} Abramowitz M and Stegun I A 1970 \textit{Handbook of mathematical functions} (New York: Dover)
\bibitem{gsk1} Green A E S, Schwartz J M and Khosravi A 1986 \textit{Phys. Rev.} A \textbf{33} 2087
\bibitem{hall} Hall R L 1992 \textit{J. Phys. A: Math. Gen.} \textbf{25} 1373
\bibitem{g1} Green A E S 1982 \textit{Phys. Rev.} A \textbf{26} 1759
\bibitem{lag} Semay C, Baye D, Hesse M and Silvestre-Brac B 2001 \textit{Phys. Rev. E} \textbf{64} 016703 
\bibitem{gara91} Garavelli S L and Oliveira F A, \textit{Phys. Rev. Lett.} \textbf{66} 1310 (and references therein)
\bibitem{env0} Hall R L 1983 \textit{J. Math. Phys.} \textbf{24} 324; 1984 \textit{J. Math. Phys.} \textbf{25} 2708
\bibitem{hult51} Hulth\'{e}n L and Laurikainen K V 1951 \textit{Rev. Mod. Phys.} \textbf{23} 1
\end{thebibliography}
\end{document}